\documentclass[review]{elsarticle}
\usepackage{xcolor}
\usepackage{lineno,hyperref}
\usepackage{float,geometry,graphicx}
\usepackage{graphics}
\usepackage{geometry}
\usepackage{bbold}
\modulolinenumbers[5]
\usepackage[export]{adjustbox} 
\usepackage{xspace}
\journal{}

\begin{document}
\nolinenumbers
\begin{frontmatter}

\title{Jackstraw Inference for AJIVE Data Integration}
\author{Xi Yang\footnote{xiyang@ad.unc.edu}, Katherine A. Hoadley\footnote{hoadley@ad.unc.edu}, Jan Hannig\footnote{hannig@ad.unc.edu}, J. S. Marron\footnote{marron@ad.unc.edu}}
\address{The University of North Carolina at Chapel Hill}

\begin{abstract}
In the age of big data, data integration is a critical step especially in the understanding of how diverse data types work together and work separately. Among data integration methods, the Angle-Based Joint and Individual Variation Explained  (AJIVE) approach is particularly attractive because it not only studies joint behavior but also individual behavior. Typically AJIVE scores indicate important relationships between data objects, such as clusters. An important challenge is understanding which features, i.e. variables, are associated with those relationships. This challenge is addressed by the proposal of a hypothesis test for assessing statistical significance of features.  The new test is inspired by the related jackstraw method developed for Principal Component Analysis.  We use a high-dimensional muti-genomic cancer data set as our strong motivation and deep illustration of the methodology.
\end{abstract}

\begin{keyword}
Data integration; AJIVE; jackstraw; statistically significant. 
\end{keyword}

\end{frontmatter}

\section{Introduction}
\label{intro}
As modern data sets have grown large they have also grown more complex.  A canonical example is cancer genomics research.  That field made a major transformation in the direction of Big Data over 20 years ago with the advent of micro-arrays for high throughput measurement of gene expression, resulting in simultaneous quantification of the activity of tens of thousands of genes in a single tissue sample.  Since then gene expression measurements have been refined by several orders of magnitude using next generation sequencing also called RNA seq. These technologies led to many new biological discoveries, and also drove the development of many novel statistical methodologies.  

An important method for obtaining deep biological insights from such high dimensional data has been Principal Component Analysis (PCA) \cite{hotelling1933analysis}.  As noted in Section 3.1.4 of \cite{marron2021object}, PCA provides many data insights through {\it modes of variation}.  These are a sequence of orthogonal rank one approximations of the mean centered data matrix, representing maximal variation explained.  They are the products of {\it loading vectors}, representing directions of maximal variation in the data space, with {\it scores vectors}, that are the projection coefficients of each data case onto the loading vectors. {Alternatively, modes of variation can be defined using Singular Value Decomposition (SVD) of the centered data $D = U \Lambda V^\top$, where $U=(u_1,\ldots,u_r)$ and $V=(v_1,\ldots,v_r)$ are matrices comprised of $r$ orthonormal columns and $\Lambda$ is a diagonal matrix with positive entries \citep{GolubKahan1965}. Each rank~1 matrix $u_i \lambda_i v_i^\top,\ i=1,\ldots, r$ is a mode of variation.

PCA scores are very interpretable as they give a clear impression of how the high dimensional data vectors relate to each other (often called dimension reduction). Scatter plot matrix views of scores often give clear views of data set structure, e.g. finding scientifically important clusters or groups of samples with shared features (see Section 4.1 of \cite{marron2021object}).  The feature loadings are also key to interpretability as they provide insight about the features driving the population structure discovered in the score plots.  As an example, genes with major involvement in cancer subtypes (clusters) tend to have large magnitude loadings. An important question is which of these loadings are statistically significant which is not evident from PCA analysis alone.  To overcome this, the jackstraw method \cite{chung2014statistical} was developed and can viewed as a permutation version of jacknife \citep{Quenouille1949} designed for high-dimensional problems.

Despite the many useful biological discoveries made from measuring high dimensional gene expression, cancer still remains a major medical challenge, due to incredible diversity and complexity. In addition to gene expression, there are many other high-dimensional assays used in cancer that include other measurements of biological functions including mutations, copy number variation, DNA methylation, miRNAs, and protein expression.  It is important to understand how these different measurements work together (e.g. a copy number variation that amplifies a gene region leading to more copies of the gene and increased gene expression) or where a data type has independent information. This has motivated the statistical area of {\it multi-block} analysis (also referred to {\it multi-view} in the machine learning literature or {\it multi-omics} in bioinformatics).  These analyses consist of multiple data matrices (one for each data type) called blocks, which consist of data vectors coming from a {\it common} set of experimental units (e.g. tissue samples). 

An important example of multi-omics data is The Cancer Genome Atlas (TCGA) \cite{hutter2018cancer}. The present paper focuses on just two of the data types available in TCGA.  The first is Gene Expression (GE), widely known to be fundamental to cancer biology.  The second is Copy Number Variation (CNV),  which quantifies repeated replication and also deletion of chromosomal parts (that play an important role in many cancer types).  In this paper, we work with Copy Number Region (CNR), medians of CNV over chromosomal regions called cytobands. There is strong biological interest in the association between GE and CNV, which motivates our AJIVE - Jackstraw analyses.

A useful method for analyzing multi-block data is the Angle-based Joint and Individual Variation Explained (AJIVE), proposed by  \cite{feng2018angle}.  In a similar spirit to PCA, AJIVE reveals population structure through a decomposition into modes of variation.  However, AJIVE takes multi-block analysis to the next level by providing two types of such modes of variation.  The first type is {\it joint} between data blocks, in the sense of having {\it common} scores vectors for each block. Here, each data block represents a different type of data measurement made on the same set of experimental samples and therefore it is possible to have common scores, but not possible to have common loadings. The second type are modes of variation that are {\it individual} in the sense of having scores vectors that are orthogonal to the joint scores.  Both types of scores are usefully visualized to find interesting population structure. The loadings indicate which features in each data block work together (joint) or separately (individual).

The main contribution of the present paper is the adaptation of jackstraw for inference on loadings in AJIVE analysis.  In particular, it assigns statistical significance to both joint and individual loadings, which indicates the significance of the ranked feature loadings.  Basic ideas are illustrated using a Toy Example in Section \ref{toy}.  Section \ref{al} gives the algorithmic development of our novel jackstraw for AJIVE.  Application of our AJIVE based jackstraw, in the context of cancer genomics research appears in Section~\ref{tcga}.}

\subsection{Introduction to AJIVE}
\label{toy}
 Following the standard in bioinformatics, the columns of these matrices represent data vectors or different experimental units and the rows represents features sometimes called variables or traits. In particular, all data matrices have the same number of columns (experimental units) but potentially different number of rows (features). This is the transpose of what is usual in classical statistics. Using this convention, AJIVE \cite{feng2018angle} assumes the following model for each of the mean centered data blocks:
\begin{equation}\label{eq:AJIVEmodel}
 D^m = L_{J}^m \times V_{J}^\top + L_{I}^m\times V_{I}^{m \top} + E^m,\quad m=1,\ldots, M.
\end{equation}
Here the low rank matrix of joint scores $V_{J}$ is common for all the data blocks and is composed of $r_J$ orthonormal columns. The low rank matrices
of individual scores $V_{I}^m$ are composed of $r_{I}^m$ orthonormal columns that are orthogonal to $V_J$. Notice that opposite to the usual PCA convention, to give the AJIVE analysis scale free properties, it is the scores that are normalized to have norm 1 and the loadings (columns of $L_J^m$ and $L_I^m$) reflect the variation quantified by singular values.
The matrix $V_{J}$ is determined only up to a left multiplication by an orthogonal matrix. The matrices $V_{I}^m$ are chosen so that, just as in SVD, the columns of $L_{I}^m$ are orthogonal and decreasing in norm. The $E^m$  are full rank residual noise matrices.

The AJIVE modes of variation are obtained as outer products of the corresponding columns of the loading and score matrices, i.e., if $L_{i,J}^m$ and $V_{i,J}$, $i=1,\ldots,r_J$, are the $i$th column of $L_{j}^m$ and $V_J$ respectively, the
$i$th joint mode of variation for block $m$ is the rank~1 matrix $L_{i,J}^m \times V_{i,J}^\top$. Analogously, the $j$th individual mode of variation is the rank~1 matrix $L_{j,I}^m \times V_{j,I}^{m\top}$, $j=1,\ldots, r_I^m$.

The AJIVE estimator of $V_{J}$ is called the {\em Common Normalized Scores} (CNS) matrix, and the estimators of $V_{I}^m$ are called the {\em Block Specific Scores} (BSS). Similar to PCA scores, the CNS and BSS indicate the relationships between the data vectors. Again an important difference from traditional PCA is that the columns of CNS and BSS are normalized to be orthonormal vectors, hence they are also direction vectors in $\mathbb{R}^n$. This is because AJIVE inference is based on studying uncertainty in the angle distribution of the scores.
In particular, the joint space, i.e. subspace of $\mathbb{R}^n$ spanned by the columns of CNS, defines the estimated joint variation of all data blocks among the experimental subjects. Similarly the subspace of $\mathbb{R}^n$ spanned by columns of each of the BSS defines the estimated individual variation.

Estimated modes of variation are obtained from the estimated scores and loadings matrices as outer products of the corresponding columns. The joint and individual estimated modes of variation reflect how the data blocks work together and separately respectively. An important measure of the effectiveness of AJIVE in recovering these modes of variation is the angle between the estimated loadings vector and the true underlying direction.

\begin{figure}[H]
  \centering
  \makebox[\textwidth][c]{\includegraphics[width=1\textwidth]{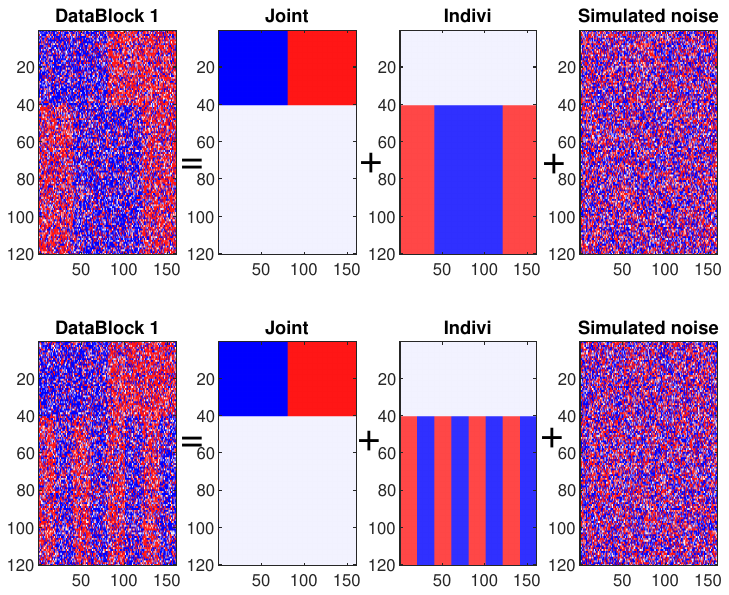}}
  \caption{This figure illustrates how we construct the toy example. The top and bottom panels show DataBlocks 1 and 2 respectively. The rows represent features/variables and columns represent cases/samples. Each data block has 120 features (rows) and 160 cases (columns). From left to right, the panels are: input to AJIVE, simulated joint, individual, and noise matrices. Red indicates values greater than 1 and blue indicates values less than -1, in-between values are shown with less color intensity and white is zero. The simulated individual signals are weaker than the joint signal by 20\% in each data block as indicated by the slightly  paler colors.}
  \label{AJIVEreal}
\end{figure}

Using a simple Gaussian simulation, we illustrate how AJIVE identifies modes of variation from multiple data blocks that is the basis of the inference developed here (Figure~\ref{AJIVEreal}). This toy example is constructed as in (\ref{eq:AJIVEmodel}) across two data blocks to visibly demonstrate the difference between joint and individual modes of variation. In real data, this is not as clear as in this toy example. Figure~\ref{AJIVEreal} shows a number of panels. Each is a heat map view of a data matrix whose entries are coded by colors. The color red is coding values that are greater than 1 and blue is coding smaller than -1. White codes zero and in-between values are lighter shades of red or blue. As discussed above, the columns of these matrices represent data vectors or different experimental units. The rows of the matrices represents features sometimes called variables or traits.  This toy example has two data blocks: Datablock 1 (shown on the top) and Datablock 2 (bottom). Both have 120 features (rows) and 160 cases (columns).  The input data matrices of the toy sample in the far-left panels are the sum of the other three matrices representing joint variation, individual variation, and noise respectively  shown in the other panels. The second panels on the left are rank-1 joint matrices. The first 40 rows in the top (Datablock 1) have a similar pattern to the first 40 rows in the bottom, representing important joint features which should be flagged as significant by jackstraw and the remaining rows should be insignificant regarding the joint modes of variation. The third panels on the left are rank-1 individual matrices and the top and bottom panels are very different. They have a relatively weak signal characterized by a minimum value of -0.8 and a maximum of 0.8, as indicated by the paler red and blue colors. The individual jackstraw should label the bottom 80 rows as significant.  The far-right panels show simulated independent Gaussian noise with mean 0 and variance 2. 

The different joint/individual signal strengths in the toy example in Figure \ref{AJIVEreal} are designed to show how AJIVE recovers signals that are poorly recovered by PCA. This example will also highlight the fact that this richer information improves the statistical power of AJIVE jackstraw tests relative to PCA.

\begin{figure}[H]
  \centering
  \makebox[\textwidth][c]{\includegraphics[width=1.2\textwidth]{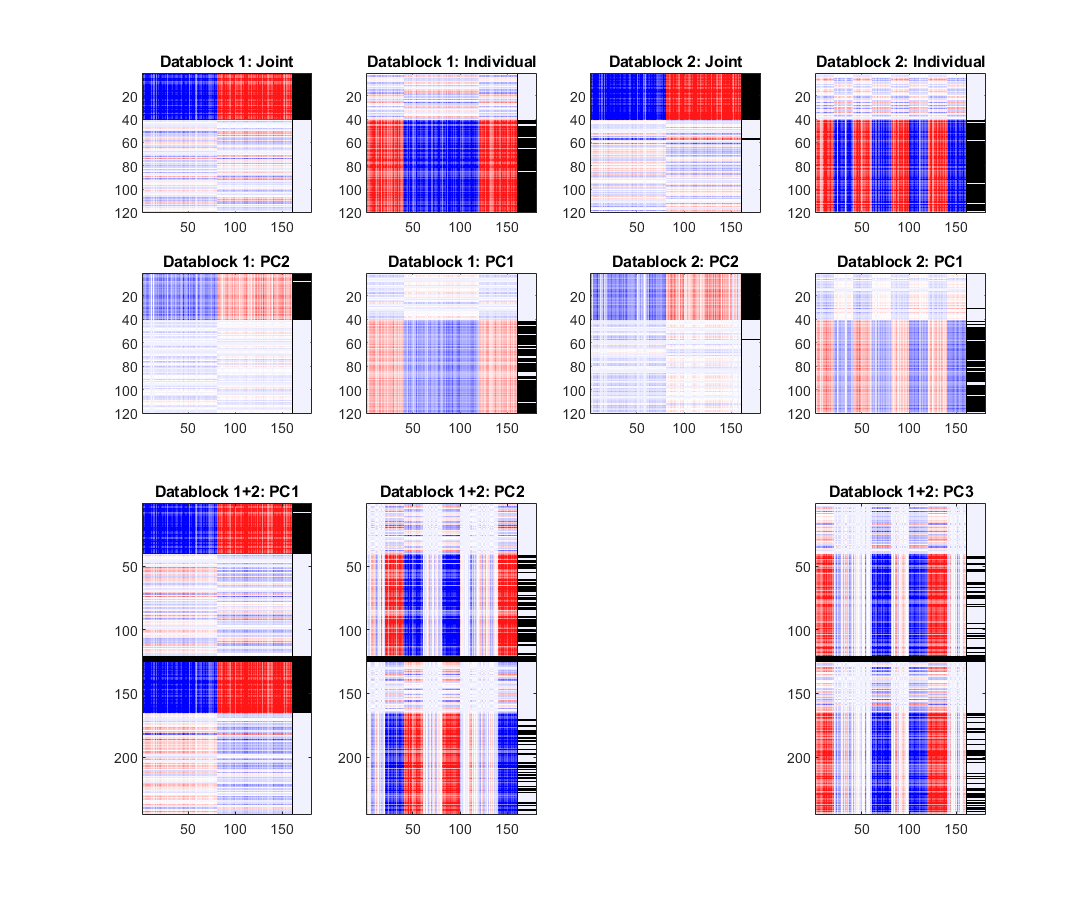}}
  \caption{Comparison between the AJIVE-jackstraw (top row), PCA-jackstraw applied to each data matrix separate (middle row), and PCA-jackstraw applied to a concatenated data matrix. Each heatmap represents the data projected in the corresponding direction indicated in the title with the same color code as Figure \ref{AJIVEreal}. The black and white columns on the right of each heatmap indicate the jackstraw significant features. The AJIVE-jackstraw provides more accurate results as indicated by the stronger red blue colors and more accurate black and white bars.}
  \label{Jackpca}
\end{figure}

As noted above the jackstraw method \cite{chung2014statistical} for statistical inference on PCA loadings has inspired our approach to AJIVE inference. Technical details are given in Section \ref{al}. To demonstrate the improvement of jackstraw with AJIVE over PCA, we applied jackstraw to three different sets of data.  1) AJIVE extracted rank-1 Joint and Individual datasets (Figure \ref{Jackpca}, top row), 2) PCA modes of variation on Datablocks 1 and 2 separately (Figure \ref{Jackpca}, middle row) ordered by best correspondence to the row above, and 3) PCA on the concatenated datablocks 1 and 2 (Figure \ref{Jackpca}, bottom row). Jackstraw significance is indicated by horizontal black lines on the right of each heatmap matrix (using the same color code as Figure \ref{AJIVEreal}).

The first major take way from this analysis, was that stronger signal, indicated through deeper red and blue colors, are found in AJIVE compared to PCA (Figure \ref{Jackpca}, row 1 versus row 2).  Concatenation of the datablocks did improve ability of PCA to find joint signal but was not able to resolve individual modes of variation similar to AJIVE (Figure \ref{Jackpca}, row 3 versus row 1).  The second major take away is that the ability for jackstraw to identify the significant joint and individual components.  This is observed by the higher number of joint (top 40 rows) and individual (bottom 80 rows) identified as significant by jackstraw in AJIVE compared to PCA.

\section{Jackstraw Inference}
\label{al}
Recall the definitions of CNS and BSS and corresponding notation at the beginning of Section~\ref{toy}. Let $V_{n\times 1}$ be one column of either $CNS_{n\times r_J}$, i.e. scores for one joint mode of variation, or $BSS_{n\times r^m_I}$, i.e., scores for one individual mode of variation.

For one datablock $m=1,..., M$, jackstraw inference is based on either joint or individual loading vectors. The corresponding loading vector (one column in the estimated loading matrix) can be computed as:
\[L_{d_m\times 1}=D_{d_m\times n}\times V_{n\times 1}.\]
This can also be viewed as the least squares solution of the following linear regression problem:
\begin{equation}\label{eq:jackstraw}
(D_{d_m\times n})^T=V_{n\times 1}\times(L_{d_m\times 1})^T+(E_{d_m\times n})^T,
\end{equation}
Note that Equation \ref{eq:jackstraw} contains $d_m$ separate simple linear regressions without an intercept in the following sense:
\begin{itemize}
\item Each column of $(D_{d_m\times n})^T$ is the response $Y$ in the linear regression. It is centered before AJIVE so no intercept needs to be considered.
\item $V_{n\times 1}$ represents the predictor of the linear regression. 
\item The corresponding entry of $L_{d_m\times 1}$ is the unknown simple linear regression coefficient, i.e. $\beta$ in simple linear regression.
\item Rows of $E_{d_m\times n}$ are the residual error terms.
\end{itemize}

The aim of the proposed jackstraw test is to find which entries of the loadings vector $L_{d_m\times 1}$
are statistically significant. The following hypothesis test structure forms the basis of our jackstraw inference. Let $L_{d_m\times 1}=[l_1,..., l_{d_m}]^T$. For each $j=1, ..., d_m$, we consider a hypothesis test:
$$H_0: l_j=0,\ \ \  H_1: l_j\ne 0.$$
For each of these tests, we define the test statistic as
$$F_j=\frac{(SSE_0-SSE_1)/1}{SSE_1/(n-1)},$$
where $SSE_0$ and $SSE_1$ are the residual sums of squares under $H_0$ and $H_1$.

Unlike in a classical regression model, we do not expect $F_j$ to follow an F distribution due to the data not necessarily being normal. For each entry $l_j$, jackstraw could be viewed as a regular permutation test, where entries of the corresponding row of $D_{d_m\times n}$, i.e. response in the corresponding simple linear regression, are permuted. Therefore we proposed to estimate the distribution of $F_j$ under the null hypothesis using jackstraw. The proposed algorithm generates the simulated null distribution by randomly choosing $K$ rows and within each row randomly permuting the observations. To do this correctly, one needs to rerun an AJIVE to re-estimate the $V$ matrices. This is computationally expensive and therefore we recommend selecting $K$ rows,  permuting those labels then rerunning AJIVE. After that, we do the simple linear regressions corresponding to the rows of $D_{d_m\times n}$. To generate the null distributions this is repeated $S$ times. In high dimensional settings we expect some of the rows of the data matrix to be very correlated. Therefore, we only want to permute a small number of rows perhaps even $K=1$.

Thus we have $S\times K$ test statistics simulated under the null hypothesis. The parameters $K$ and $S$ are chosen as follows:
\begin{itemize}
\item $K$ is the number of rows selected in one step, which should be much less than $d_m$, such as 1, 10, or 100. 
\item $S$ is the number of times the permutation step is repeated, which should be at least 10 times $d_m$.
\end{itemize}
The key algorithmic steps are:
 \begin{enumerate}
    \item The test statistics, $F_js$, are obtained by fitting a linear regression with the rows of mean-centered AJIVE input data as the response and with columns of the selected CNS or BSS of interest as the predictors. 
    \item  Randomly select $K$ rows of the original data matrix and permute the observations within each row and recalculate the AJIVE CNS and BSS. Then recalculate the test statistics as in Step 1 using the permuted data matrix and the same columns as in Step 1 of the new CNSs or BSSs to get K samples from the simulated null distribution. Then Repeat $S$ times to generate $S\times K$ samples from the null distribution $F_{null}^b, b=1, ..., S\times K$.
    \item  Calculate an empirical p-value: $p_j=\frac{\sum_{b=1}^{S\times K}I(F_j<=F_{null}^b)}{S\times K}$ for each observed test statistic $F_j, j=1, ..., d_m$, where $I$ denotes the indicator function.
    
   \item Adjust p-values for multiple testing. We recommend using the Bonferroni adjustment \cite{bonferroni1936teoria} for the level of the test, i.e. the regular p-value times the number of tests (in this case the number of features $d_m$) as the adjusted p-value. A reasonable alternative would be to consider the false discovery rate \cite{benjamini1995controlling}.
\end{enumerate}

As stated in \cite{chung2014statistical}, smaller $K$ (in Step 2) gives more precise estimation, and larger $K$ gives faster calculation. The analyses in this paper use $K=1$ and parallel computation for maximal precision per unit time.

\subsection{Approximate Algorithm for Faster Computations}
Each of the $S$ resamples requires one round of the AJIVE algorithm. 
When the data matrices are large recomputing AJIVE $S$ times could be very computationally intensive. In many situation, the rows are highly correlated and individual features tend not to be particularly influential such as the data in Section \ref{tcga}.
Thus permuting a small number $K$ out of a large number $d_m$ of rows will lead to minor changes in the AJIVE results. This is especially true in the case of a joint space estimated from multiple data blocks. In that case, one can save a large amount of computational effort by just using the original CNS rather than recomputing AJIVE $S$ times. Note this results in a classical permutation test.

We next demonstrate the effects of not recomputing AJIVE for each of the $S$ replications of jackstraw using TCGA breast cancer data, which will be analyzed in detail in Section \ref{tcga}. As noted above, we have 2 data blocks (data matrices): 
CNV that includes as features (rows) CNR, and
GE that includes as features normalized log of gene expression.

After AJIVE inference, we identify 3 joint components and we proceed to apply jackstraw to find the significant genes/CNRs corresponding to each of the 3 joint components. As an investigation of the Monte Carlo variation, we run the 2 algorithms (full and approximate) 2 times using 2 different seeds, denoted as seed1 and seed2. For convenience of notation, let `App' denote the approximate algorithm and `Full' denote full algorithm. In Table \ref{number}, we will report the number of significant CNRs/genes. After a careful comparison, more than 95\% of the significant CNRs/genes stay significant if we use different algorithms or random permutations. In each row, the subset corresponding to a smaller number of significant features is always included in the subset corresponding to the larger number. Thus, jackstraw analysis of TCGA breast cancer data is not sensitive to random permutations or algorithms.
\begin{table}[H]
\centering
\caption{Exploration of consistency of significant features. Rows correspond to data blocks and joint component numbers. Columns indicate the number of significant features for each of the two algorithms,  approximate (App) and full. Also explored is the impact of the particular realization of the random permutations labeled 1 and 2 in the 3rd to 6th columns. The last column is the total number of GE and CNV features. This suggests that randomization only has a minor influence. Recall, CNV is an abbreviation for Copy Number Variation, GE for Gene Expression, and CNR for Copy Number Region.\\}
\begin{tabular}{lllllll}
\hline
Datablock&Joint Component & App-1 & Full-1 & App-2 & Full-2 & Total number\\ \hline
GE&1 & 11288 & 11450 & 11450 & 11051 & 20249\\ \hline
GE&2 & 7058 & 7765 & 7859 & 6918 &20249\\ \hline
GE&3 & 6274 & 6510 & 6455 & 6288 &20249\\ \hline
CNV&1 & 526 & 557 &526  &  537& 806\\ \hline
CNV&2 & 247 & 197 & 276 & 276 &806\\ \hline
CNV&3 & 302 & 235 & 240 & 247 &806\\ \hline
\end{tabular}
\label{number}
\end{table}
While the algorithms gave similar results, the approximate algorithm took about 30 mins and the full algorithm takes about $30 \times d_m=30\times 20249=607470$ mins for GE without parallel computation. With access to parallel computation, the full algorithm can be much faster. In practice, we recommend making a careful choice between the 2 algorithms based on whether any features are expected to be dominant.
\subsection{Jackstraw Diagnostic Graph}
\begin{figure}[H]
  \centering
  \makebox[\textwidth][c]{\includegraphics[width=1.1\textwidth]{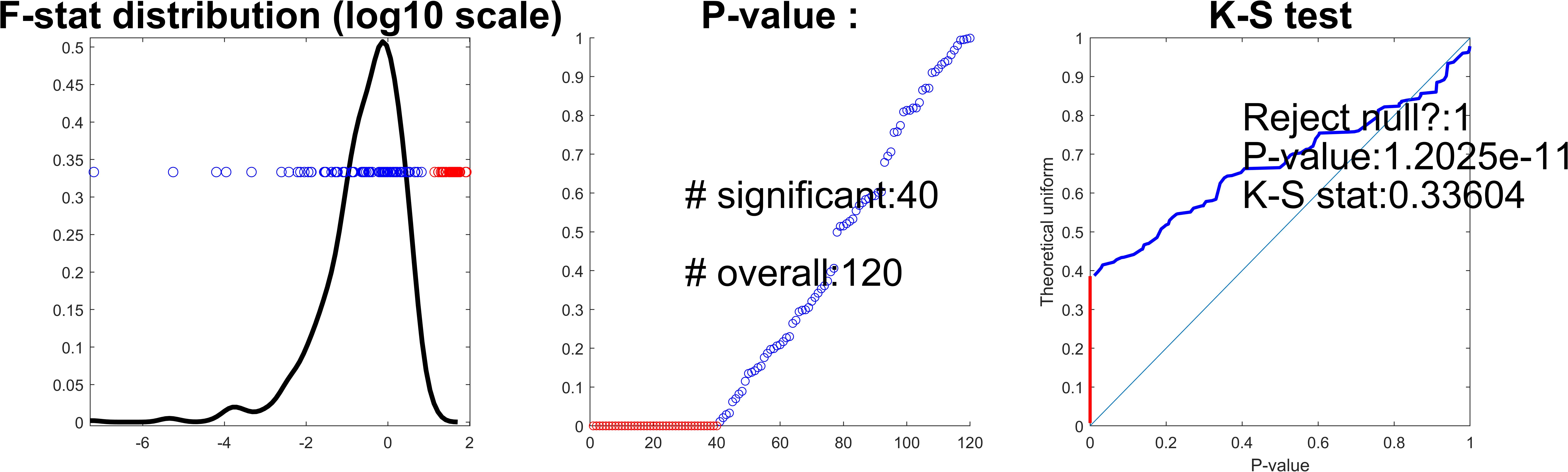}}
  \caption{Jackstraw diagnostic graphs using the simulated toy example in Figure \ref{AJIVEreal}. Left panel: the black curve is a kernel density estimate of the null distribution and the red (significant) and blue (not significant) dots are the observed F statistics for all genes (both on $log_{10}$ scale). Middle panel: the sorted p-values, one for each feature. Right panel: Kolmogorov-Smirnov test of the uniformity of the p-value. This diagnostic graph shows that the first joint component is statistically significant and there are 40 statistically significant features, which are the main drivers of this joint component.}
  \label{diag}
\end{figure}

Jackstraw typically involves a very large number of individual hypothesis tests. We recommend three diagnostic graphs (kernal density estimates, jackstraw p-values, and Kolmogorov-Smirnov tests) as useful methods for quickly summarizing the results of the numerous individual hypothesis tests such as those shown in  Figure \ref{diag} of the toy example in Section \ref{toy}. In the left panel, the black curve is a kernel density estimate visualization of the $F_{null}^b$ distribution (on the $log_{10}$ scale) calculated in Step 2 of the algorithm in Section \ref{al}. The red (significant) and blue (not significant) dots are the observed F statistics: $F_j$ calculated in Step 1. Significance is at the level of 0.05 using the Bonferroni correction. The middle panel shows the sorted jackstraw p-values, computed in Step 3, for all the genes. Out of 120 features, we find 40 significant features in the joint component. The right panel shows the results of a Kolmogorov-Smirnov test (K-S test)\cite{walsh1963bounded} for whether the first component overall has useful information. This tests the uniformity of the distribution of these p-values. In this panel, the colored curve is relatively far from the 45-degree line and the K-S p-value is $1.2\times e^{-11}$.
\section{TCGA Breast Cancer Data}
\label{tcga}
As noted in the introduction, The Cancer Genome Atlas (TCGA) \cite{hutter2018cancer} is a multicenter effort to
generate multiple different data types for a large cohort of patients to comprehensively characterize cancer. TCGA contains many cancer types and several different data platforms. We are particularly interested in the breast cancer \cite{ciriello2015comprehensive} cohort, where prior work has identified molecular subtypes of breast cancer, based on gene expression \cite{perou2000molecular}, \cite{sorlie2001gene}. Therefore, this data set is an excellent case to study the application of jackstraw on the integration of multiple data platforms. The 1038 patients are classified into 4 subtypes (185 Basal cases, 81 Her2, 556 LumA, and 216 LumB). Of particular interest are the relationships between GE and CNV and how those relationships vary by subtype. Hence we do a three block AJIVE analysis here: GE ($d_1=20249$), CNV ($d_2=806$), and subtype ($d_3=4$) and $n=1038$. Inclusion of the third block provides a {\it supervised} version of AJIVE, which enables careful focusing on the subtypes. The indicator (subtype) matrix has a column corresponding to each of the 1038 patients. Each column of the matrix has four entries, which indicate the four subtypes using a one in the appropriate row and zeroes for the other entries. 

We use the $log_2$ count GE values. To construct the CNV data, we use the $log_2$ ratio per gene values from GISTIC downloaded from TCGA FireBrowse website. All data blocks are feature-centered as the input of AJIVE. Hence, the sign of all features is relative to the average. 

 A low-rank approximation is used in the first step of AJIVE for the first two input matrices. Using AJIVE diagnostic plots we established low ranks of 77 for CNV, 70 for GE and 3 for the subtype indicator matrix. This resulted in a rank 3 joint space i.e. 3 joint components overall, 74 and 67 individual components for CNV and GE respectively. In the following sections, we investigate the statistical significance of the loadings of the AJIVE joint and individual spaces using jackstraw.
\subsection{Joint Space}
\begin{figure}[H]
  \centering
    \makebox[\textwidth][c]{\includegraphics[width=1\textwidth]{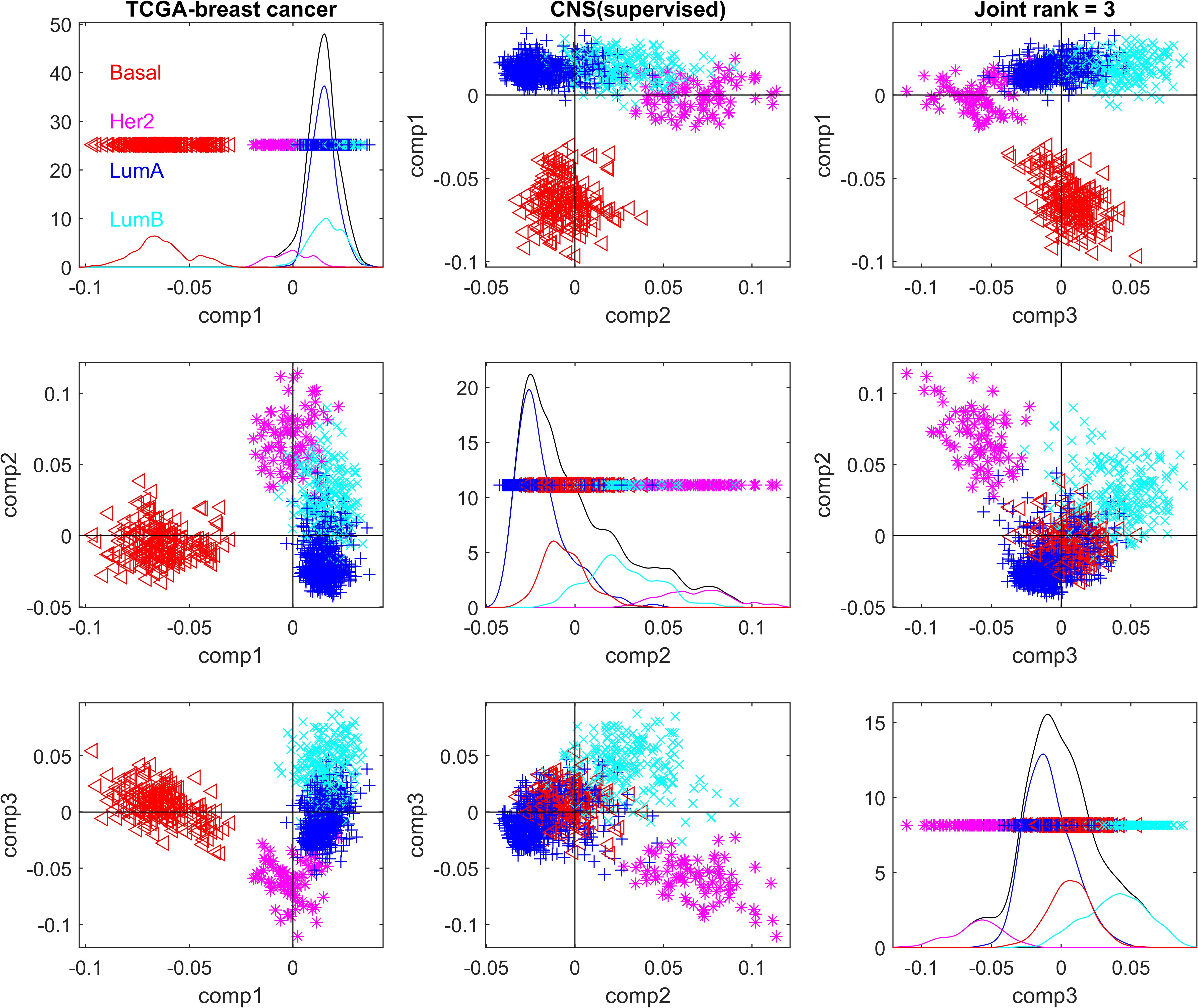}}
  \caption{Common normalized scores scatter plot for joined component of supervised AJIVE. Each column of CNS corresponds to one of the joint component, i.e. scores. The diagonal panels are jitter plots and kernel density estimates of the univariate distribution of the entries of the corresponding CNS vector. The off-diagonal panels are related scatter plots giving a pairwise comparison of the entries of the score vectors. The subtypes are indicated as follows: Basal: red; Her2: magenta; LumA: blue; LumB: cyan. This figure shows that the joint variation reflects strong cancer subtype separation.}
  \label{cns}
\end{figure}

\begin{figure}[H]
  \centering
  \makebox[\textwidth][c]{\includegraphics[width=1\textwidth]{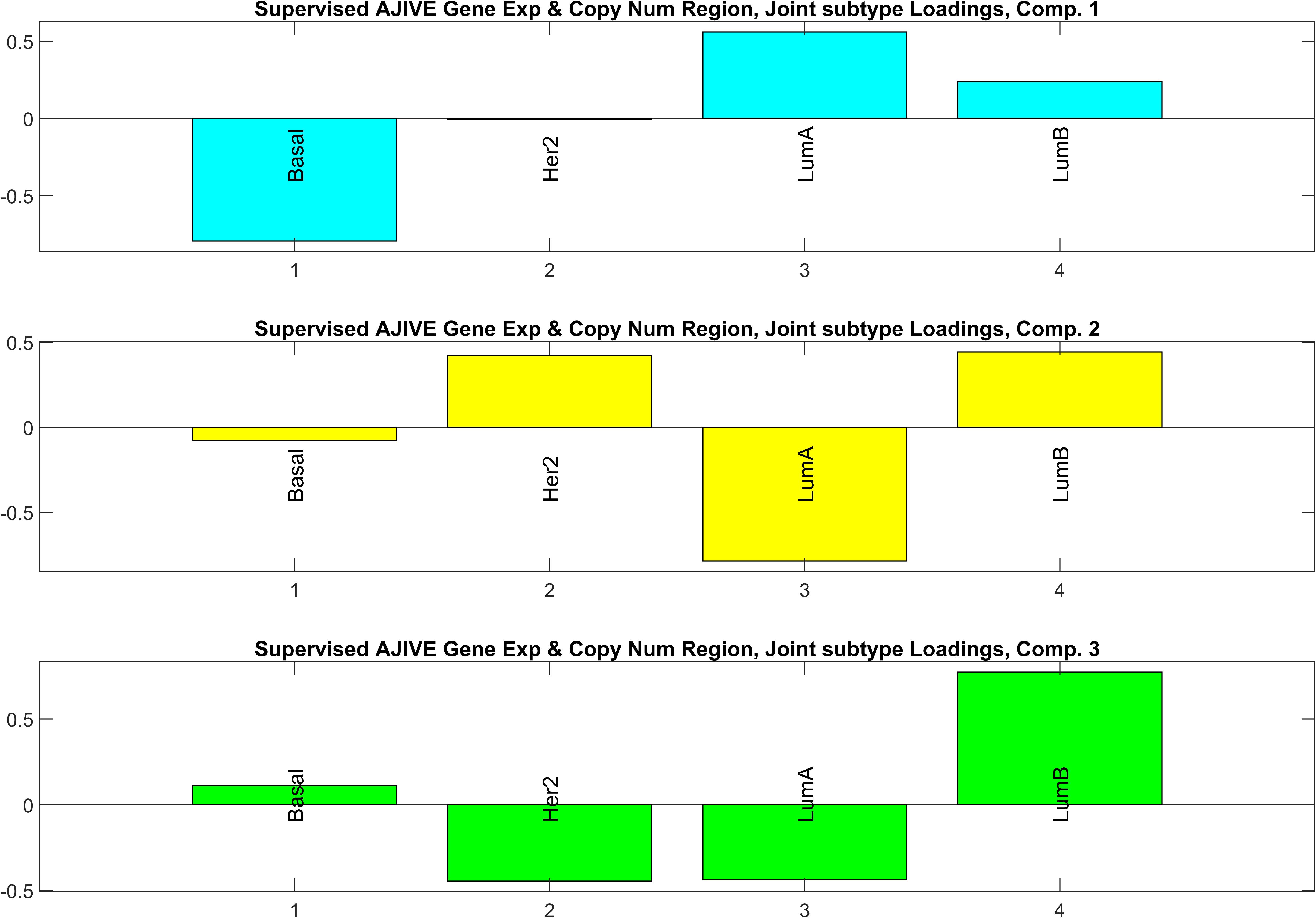}}
  \caption{CNS loadings of the subtype matrix indicating the impact of each subtype on the respective mode of variation. This figure shows the subtype contrasts that drive each joint component.}
  \label{barbar}
\end{figure}

The following discussion demonstrates the kind of information that can be learned from the Common Normalized Scores (CNS). Figure \ref{cns} shows a scatter plot matrix view of the CNS, which indicates how the cancer patients relate to each other in terms of GE and CNV. This shows a strong visual separation of the subtypes, that is a consequence of subtypes playing an important role together in GE and CNV, as well as supervision using the third data block. The CNS ($V_{n\times1}$ in (\ref{eq:jackstraw})) is a 1038 by 3 matrix because of 3 joint components and 1038 cases/patients. The plots on the diagonal are one-dimensional views of the scores, columns of the CNS matrix. The subtypes are illustrated with subplot kernel density estimates (KDE). Scores are shown using jitter plots (i.e. random heights) to visually separate them in the one-dimensional plots. The overall KDEs are shown in black and the subtype KDEs are shown in subtype colors. The area under the curve of each colored KDE is proportional to its abundance so that the sum of the areas under the colored curves equals 1, the area under the black curve. The off-diagonal plots are scatter plots, i.e. two-dimensional projections of the data onto the planes generated by the corresponding CNS vectors. The first joint component  strongly separates the Basal subtypes (red). Joint components 2 and 3 also seem to be strongly related to subtypes even though the separation is less than component 1. The second joint component  separates the three other subtypes in the order of LumA (blue), LumB (cyan), and Her2 (magenta). This is sensible because these subtypes are known to be closer together and Her2 is known to be more distinct from LumA. The third joint component separates Her2 (magenta), LumA (blue), and LumB (cyan). 
\label{J}
For a deeper understanding of this relationship between subtypes, Figure \ref{barbar} shows the subtype loading for each joint mode of variation from Figure \ref{cns}.  These loadings indicate the impact of each subtype on the mode of variation. This gives another interpretation of the 3 joint components:
\begin{enumerate}
    \item contrast of Basal vs the rest; contains little Her2 information; 
    \item contrast of (Her2 \& LumB) vs LumA; contains little Basal information; 
    \item contrast of LumB vs (Her2 \& LumA); contains little Basal information.
\end{enumerate}

While AJIVE analysis indicates distinct joint spaces, we implement the jackstraw method to focus on the driving set of significant features for each joint component. The column App-1 in Table \ref{number} shows the number of jackstraw significant genes and CNRs for each joint component, using the approximate algorithm. For joint component 1, the strong differences between Basal-like subtype and non-Basals in Figures \ref{cns} and \ref{barbar} appear in column App-1 in Table \ref{number} as a large number of significant genes. This may be surprising to statisticians who have approached gene selection via sparsity methods, which has a goal of very few significant genes. However, it clearly underlines the biological complexity of this cancer. The differences within the non-Basal subtypes are weaker, which reflects itself in a smaller number of significant genes.

Our AJIVE version of jackstraw allows us to focus on a set of statistically significant features, giving insight into how GE and CNV work together. In cancer, one important mechanism is copy number variation that can influence expression by increasing or decreasing the numbers of copies per gene. Hence significant copy number variation is expected to lead to significant gene variation. We want to see if genes that are significant in joint space are joint because the copy number is directly regulating expression, however, it's important to note that there can be many genes located in one CNR.  We are interested in the \textit{cis effects} where the copy number variation directly regulates the gene expression. As shown in the second row of Table \ref{70}, almost all significant CNRs contain some significant genes. This is consistent with GE being influenced by its local copy number. 
However, as shown in the third row of Table \ref{70}, not all significant genes in joint space have a dependency on significant CNRs. There are other mechanisms for GE regulation that are still important and show up in the joint space without having a direct dependency on the CNRs. These may be related to a \textit{trans effect} (non-local regulation) or to presence of strongly correlated genes in the same pathway. 

\begin{table}[!t]
\centering
\caption{Percentage of overlapped significant genes and CNRs for each joint component. Most CNRs contain significant genes, but only some significant genes are located in significant CNRs.}
\begin{tabular}{llll}
Overlapped Joint Significant \% & comp 1 & comp 2 & comp 3 \\
\hline
Significant CNRs containing significant genes (\%)  & 95.27   & 94.57   & 91.67  \\
Significant genes located in significant CNRs (\%) & 69.17 & 39.68  & 36.27 
\end{tabular}

\label{70}
\end{table}

\begin{figure}[!t]
  \centering
    \makebox[\textwidth][c]{\includegraphics[width=1\textwidth]{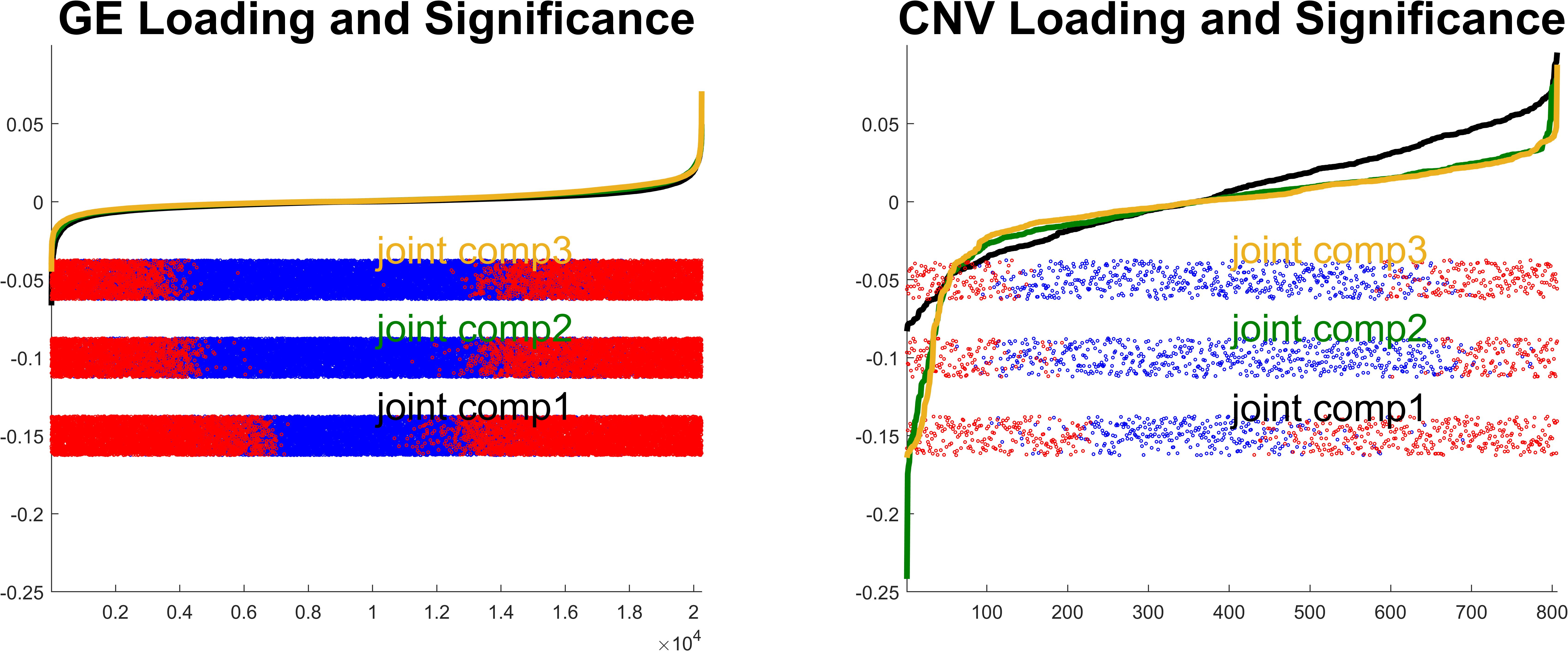}}
  \caption{The colored curves are formed by the sorted loadings of genes/copy number regions. If the corresponding feature is significant, a red dot is placed in the central bands (jitter plots), otherwise, a blue dot is placed. Features with large loadings are more likely to be selected as significant. Distributions of loadings of CNR for joint components 2 and 3 are dramatically different than what is observed in component 1.}
  \label{over}
\end{figure}

We consider how the jackstraw significance relates to the gene/CNR loadings in Figure \ref{over}.Each panel contains colored curves which show the sorted loadings for each gene/CNR. The x-axis represents the full set of the sorted features. The y-axis shows the corresponding loadings. Recall loading vectors have length one, so the sum of the squared entries must be one. In each panel, the curves show joint components 1 (black),  2 (green), and 3 (yellow). Each variable is  plotted using jitter plots (i.e. random heights) for joint components 1,2,3. Jackstraw significant genes or CNRs are identified with a red dot and non-significant features with a blue dot. The GE loadings are very similar across the three curves in the left panel, yet the CNV loadings are substantially different in the right panel. In particular, CNV has very few CNRs with very negative loadings. Relatively speaking, the signal and gene expression is spread over more genes, so there are fewer loadings that are much different from zero.   There are 20249 genes and only 806 CNRs, thus the jitter plots on the left are much denser than those on the right. In both panels, joint component 1 has more significant features than joint components 2 and 3. This is consistent with the fact that joint component 1 has the most shared variation and is driven mostly by Basal subtype information, which is known to be associated with large genomic changes \cite{hoadley2018cell}. Generally, features with large loadings tend to be more significant. 
Comparing these significant genes with prior work \cite{weigman2012basal} shows the jackstraw analysis has picked up biologically important regions previously identified in these types of cancers. Detailed analysis is shown in Section 4.3 of Xi Yang's dissertation \cite{yang2021machine}.

\begin{figure}[H]
  \centering
  \makebox[\textwidth][c]{\includegraphics[width=1.1\textwidth]{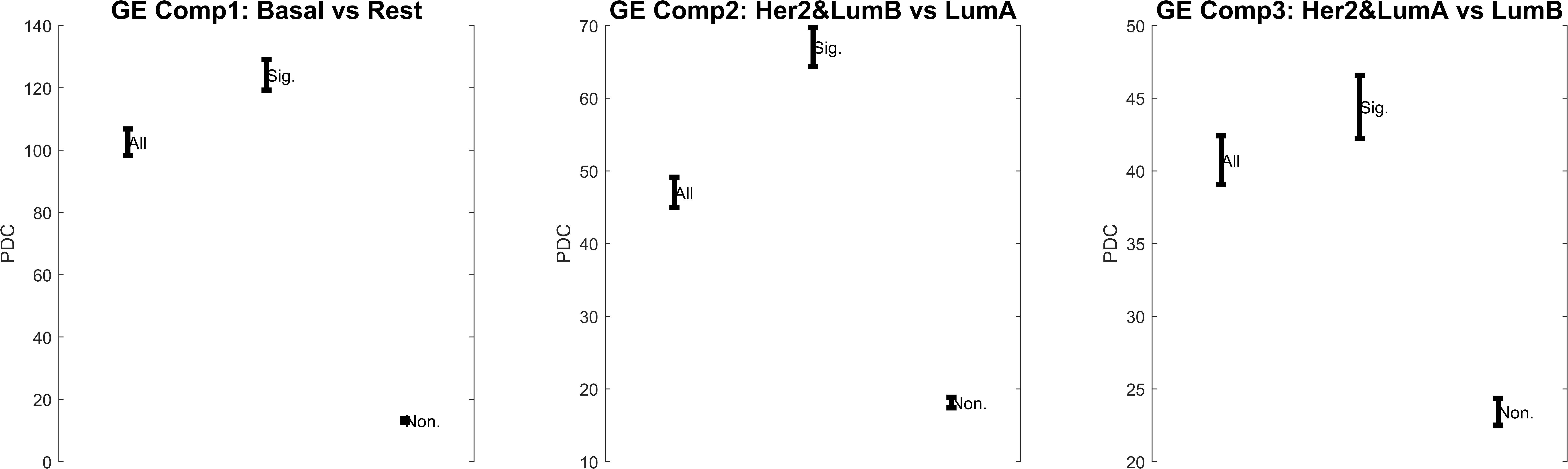}}
  \caption{DiProPerm Test PDC confidence intervals of GE expression. Left: the test of Basal vs rest, middle: (Her2\&LumB) vs LumA, right: (Her2\&LumA) vs LumB. In each panel, each error bar is a confidence interval of the PDC of the given test based on: all features (left), significant features (middle), non-significant features (right). The AJIVE-jackstraw significant features give a higher PDC verifying the stronger signal level.}
  \label{ci}
\end{figure}

Next we investigate feature selection using jackstraw. In particular, we study classification by comparing the results based on each of jackstraw significant features, jackstraw non-significant features and all features. Significance is assessed using the Direction-Projection-Permutation high dimensional hypothesis test (DiProPerm) \cite{wei2016direction}, \cite{yang2021visual} to quantify subtype separations based on different feature sets. Test results are on the scale of Population Difference Criterion (PDC) as defined in \cite{yang2021visual}. PDC works on the scale of Gaussian Z-score, so larger than 2 gives statistical significance with larger values reflecting stronger significance. Error bars are included to reflect the simulated variation of the PDC using balanced permutations as explained in Section 3.2 in \cite{yang2021visual}. Each error bar in Figure \ref{ci} shows the confidence interval of the PDC of each test, where a large PDC indicates more significance.

In each panel of Figure \ref{ci}, we quantify the amount of separation between subtypes and corresponding components indicated in the titles using 
each of the three feature subsets as the input GE data: 

\begin{itemize}
\item[1] All 20249 features (shown as `All' in the left)
\item[2] Jackstraw significant features for each joint component (shown as `Sig.' in the middle)
\item[3] Non-significant features for each joint component (shown as `Non.sig' in the right)
\end{itemize}

The comparator groups in Figure \ref{ci} were taken from the relationships observed in Figure \ref{barbar}.  Comp1 corresponds to Basal vs Rest, comp2 corresponds to (Her2\&LumB) vs LumA, and comp 3 corresponds to (Her2\&LumA) vs LumB. For all 3 panels, the tests using all features have very strong PDCs, indicating strong separations of the subtypes. When we focus on significant features only, the separations become even stronger as indicated by larger PDCs demonstrating the value of focusing on the jackstraw significance. However, the non-significant features have relatively small PDCs, but still retain some signal (PDCs greater than 2). The confidence intervals show that all of these differences are statistically significant relative to the natural permutation variation. In summary, the jackstraw significant feature sets provide the strongest subtype distinction.

\subsection{Individual space}
\label{I}
As we have seen AJIVE is very effective at finding the joint structure between all three data blocks: GE, CNV, and subtype. Next we discuss the individual information left for each data block. The DiProPerm investigation of the individual GE did not find significant subtype information, PDC of 0.59 (note less than 2). The individual CNV did reveal a significant difference between LumA and LumB, i.e. PDC of 12, which is not joint with GE. This was surprising because class labels were determined by gene expression, which is expected to have information not contained in CNV. In particular, further investigation of the individual space show that there remains informative GE-independent CNRs that can differentiate LumA and LumB. Detailed analysis can be found in Section 4.3 of Xi Yang's dissertation \cite{yang2021machine}.

\section{Discussion}
In this paper we propose a novel combination of jackstraw and AJIVE to bring needed inference. The new methodology gives more precise statistical inference about the relationship between gene expression and copy number. In particular, we found:
 \begin{itemize}
  \item The toy example demonstrated the value of AJIVE jackstraw  relative to PCA analyses. 
    \item Jackstraw analysis has picked up biologically important regions previously identified for these types of cancers.
    \item A large number of genes/copy number regions are related to the very substantial differences between Basal and non-Basals. 
    \item The smaller number of significant genes/copy number regions reflect the weaker difference within the non-Basal subtypes.
    \item Focusing on significant features (i.e., gene/copy number regions) provides a stronger subtype distinction.
\end{itemize}

\section{Acknowledgements}
This research has been partially supported by the National Science Foundation under Grant No. IIS-1633074, DMS-1916115 and DMS-211340.
\bibliography{mybibfile}
\end{document}